\documentclass[%
 reprint,
 amsmath,amssymb,
 aps,
]{revtex4-1}

\usepackage{graphicx}
\usepackage{dcolumn}
\usepackage{bm}

\begin{document}

\title{Dark matter and bubble  nucleation in old neutron stars}

\author{A. Herrero$^1$}
\author{M. A. P\'erez-Garc\'ia$^1$}%
 \email{mperezga@usal.es} 
 \author{J. Silk$^{2,3,4}$}%
 \email{silk@iap.fr}%
\author{C. Albertus$^1$}%
 \email{albertus@usal.es}

\affiliation{$^1$Department of Fundamental Physics, University of Salamanca, Plaza de la Merced S/N E-37008, Salamanca, Spain}%

\affiliation{$^2$Institut d'Astrophysique,  UMR 7095 CNRS, Sorbonne Universit\'e, 98bis Blvd Arago, 75014 Paris, France}
\affiliation{$^3$Department of Physics and Astronomy, The Johns Hopkins University, Baltimore MD 21218, USA}
\affiliation{$^4$Beecroft Institute of Particle Astrophysics and Cosmology, Department of Physics, University of Oxford, Oxford OX1 3RH, UK}

\date{\today}
\begin{abstract}
We study the probability for nucleation of quark matter droplets  in the dense cold cores of old neutron stars induced by the presence of a self-annihilating dark matter component, $\chi$. Using a parameterized form of the equation of state for hadronic and quark phases of ordinary matter, we explore the thermodynamical conditions under which droplet formation is facilitated by the energy injection  from $\chi$ self-annihilations. We obtain the droplet nucleation time as a function of the dark matter candidate mass, $m_\chi$. We  discuss further observational consequences.
\end{abstract}

               
\maketitle

\section{ Introduction} 
Neutron stars (NSs) are compact astrophysical stellar objects where the low temperature and high density regions of ordinary matter phase space can be explored \cite{reviewBaym}. Typically, their measured masses do not exceed a maximum value $M_*\sim 2M_\odot$ and radii $R_*\sim 11-13$ km. They are thought to be composed out of nucleons, mainly neutrons, with a little fraction of protons and possibly other heavier baryons or even more exotic components \cite{blashke} besides a leptonic fraction to keep electrical charge neutrality. Central nucleon number densities are thought to be several times that of nuclear saturation density, $n_0\sim 0.17\,\rm fm^{-3}$ and effective measured temperatures are in the range $T^\infty\sim 10^{5.3}-10^{6}$ K for old NSs  with lifetimes $\tau_{\rm NS}\gtrsim 10^4$ yr \cite{cool}. Under these conditions ordinary matter is typically degenerate since baryonic Fermi energies are of  the order of $E_{\rm F,B}\sim 30$ MeV, whereas the internal temperature drops below  $T\sim 1$ MeV $(k_{\rm B}=1)$ within $\sim 100$  seconds after the birth of the NS \cite{burrows}.

Although the usual description of the interior of these objects is based on effective nuclear degrees of freedom i.e. nucleons and mesons, other realizations based on quark constituents could indeed happen in nature. Early since the pioneering work by Bodmer and Witten \cite{Witten,Bodmer} the conjecture of the existence of a most fundamental, quark deconfined, state of matter in the NS core has remained an intriguing possibility. This has been explored in the literature, see for example \cite{Glend1,Heisemberg, Iida,Baym,Logoteta} and references therein. 

There are two principal  mechanisms  capable of triggering the deconfinement  transition. One involves the  increase of the central pressure \cite{bereziani} due to either the accretion of a small amount of hadronic matter or slowed rotation, while the other relies on temperature effects \cite{fluc1,fluc2,fluc3,fluc4,fluc5}. In \cite{limit}, the concept of a limiting conversion temperature in the  proto-hadronic
star is introduced as an indicator of the thermal energy $\Theta\sim 10-30$ MeV that can induce  nucleation, provided central densities (or stellar masses) are large enough. In brief, both rely on overcoming the hadronic potential barrier that confines quarks and tunneling out of the nucleon bag that is  a few fm in size. Microscopically, the locally deconfined two-flavoured $ud$ quark phase $(Q^*)$ first forms and later proceeds to a $\beta$-equilibrated three-flavoured $uds$ quark matter (QM). Note, however, that whether this former $ud$ phase truly decays has been questioned by recent works relying on arguments of energetic stability \cite{Holdom}.

Matter at such high densities can only be  partially tested by terrestrial experiments. Sites such as the GSI with the FAiR accelerator, BNL with RHIC, and CERN with the LHC, can use  heavy ion collisions (HIC) to produce the so-called quark-gluon plasma \cite{qgluon} consisting of a highly excited hadronic system   several fm in size and a lifetime of approximately 20
fm/c. Although temperatures in the fire-ball that is produced are initially high $T\gtrsim 80$ MeV,   HIC can test supra-saturation densities and provide tighter constraints on magnitudes such as the  high-density behavior of the symmetry energy, the tidal deformability and the equation of state (EoS) of nuclear matter itself, thus linking different areas of interest ranging from astrophysics to nuclear and particle physics \cite{hic}.

In this work, we are interested in the study of nucleation of quark droplets in a hadronic (nucleon) medium inside the NS core with a novel mechanism  mediated by dark matter (DM). This type of  matter is one of the key ingredients in our presently accepted cosmological model that remains as yet  undetected. It is found to constitute $\sim 26\%$ of our Universe. There are nowadays plenty of candidates from extensions beyond the Standard Model (SM) of particle physics that have been proposed to populate the dark sector \cite{bertone}.
Experimental searches with different strategies try to put constraints on the mass and scattering cross-section phase space $(m_\chi,\sigma_\chi)$. For example, while the weakly interacting candidates (WIMPs) have been thoroughly searched for in the last decade with null results \cite{pdg}, other candidates have lately attracted much attention, see \cite{axions}. Within standard cosmology, the present relic density can be calculated reliably if the WIMPs were in thermal and chemical equilibrium with the hot SM particles after inflation. In this same context, scanning the $m_\chi\sim 10-1000$ GeV range (we use $c=1$) has yielded some results. Direct detection generation II experiments based on nuclear recoils are currently approaching the $\sigma_\chi\sim 10^{-47}$ $\rm cm^2$ \cite{dd} close to the atmospheric/solar coherent neutrino interaction floor. In addition, they exclude some of the preferred regions arising from scans of SUSY models, see fig 26.1 in \cite{pdg} in the range $m_\chi\sim 30-60$ GeV  and $m_\chi\sim 10^2-10^3$ GeV. In  indirect searches \cite{id}, products of DM annihilation including neutrinos, gamma rays, positrons, antiprotons, and antinuclei can be detected. There are additional sources with constraining power such as those arising from 
large-scale structure of the Universe that rule out a $m_\chi\lesssim 400$ eV WIMP under the Tremaine-Gunn bound \cite{tremaine} along with   others.
It is generally believed that massive DM particles interact gravitationally and that any non-gravitational couplings are  expected to be weakly or strongly interacting in order  to at least maintain equilibrium with luminous  matter in our Universe. 

The structure of this contribution is as follows. In Sec. II we present the effective field theory approaches to describe the hadronic content of the interior of the NS with a possible quark phase. We introduce the relativistic Lifshitz-Kagan theory used to describe the induced bubble nucleation due to the presence of a component of DM inside the NS. Later, in Sec. III, we present our results discussing the thermodynamical conditions along with model details more favorable for nucleation of QM bubbles in hadronic matter. We evaluate the nucleation time taking into account the mass of the DM candidate. We discuss the different sources of uncertainty in our modelling and possible further astrophysical observable consequences.
Finally, in Sec. IV, we give our conclusions.

\section{Modeling the NS interior with a DM component}

We assume that a non-vanishing component of DM is present in the NS. In an evolved NS, this may be the result of various processes taking place during the lifetime of the star, including its progenitor phase as well. In dense DM environments such as the galactic centre, where there is a high density of neutron stars, clumps of DM of 
typical  mass corresponding to the free streaming mass,  of order $10^{-6}\rm M_\odot $ but possibly larger \cite{2006PhRvL..97c1301P}, can occasionally be accreted by NSs. This can provide a rare but  substantial enhancement of the DM component of the NS.

 As we explain below, gravitational capture and depletion processes (typically self-annihilation or decay \cite{kouvaris, perez}) modulate the DM population inside the star \cite{cermeno1,cermeno2}. Complementary constraints on DM annihilation processes from additional isotropic gamma-ray or  reionization and heating of the intergalactic gas backgrounds have been summarized in recent contributions \cite{annih}. We will consider DM that self-annihilates through reactions involving quark pairs $\chi {\bar \chi} \rightarrow q {\bar q} \rightarrow N \gamma$ producing photon final states. Note that although other channels are indeed available, for simplicity we will stick to this case in what follows but we will later discuss other possibilities. DM candidates ($\chi$) of interest to us will be those with cross sections $\sigma_{\chi N}$ scattering off nucleons (N) and masses $m_\chi$ in non-excluded regions of the currently available phase space \cite{pdg}.

 We first consider the cold NS core as a system described by the well-known relativistic lagrangian model from \cite{Walecka} consisting of a baryon sector $B=n,p$ (neutrons and protons) interacting through mesonic fields $\mathcal{M}=\sigma,\omega,\rho$ and a minimal leptonic sector $l=e$ (electrons). Besides, the usual non-linear self-interacting potential is included under the form %
 \begin{equation}
\mathcal{U}(\sigma)=\frac{1}{3}a_1 m_B(m_B-m^*_B)^3+\frac{1}{4}a_2(m_B-m^*_B)^4,
 \end{equation} 
 with $m_B$ being the bare baryon mass and $m^*_B=m_{B}-g_{\sigma B}\sigma$ the effective baryon mass in the medium. Specific values for the couplings of the baryon and meson fields as well as  particle  masses used can be found in \cite{Glendenning2}. 
 
In the cold system, baryonic density can be expressed in terms of the Fermi momentum $k_{F,B}$ for each particle component as $n_b=\sum_B \frac{k^3_{F,B}}{3\pi^2}$. Similarly, for the leptons we have $n_l=n_e= \frac{k^3_{F,l}}{3\pi^2}$. The equations for the mesonic fields in the extended system are obtained using a mean field approach so that we replace a generic field  $\phi(x)\rightarrow\langle\phi(x)\rangle=\phi$. By doing this the equations for the non-vanishing mesonic field components are obtained. For the $\omega_0\equiv \omega$ field one obtains
 \begin{equation}
 m_{\omega}^{2}\omega = \sum\limits_{B} g_{\omega B}n_{B},
  \end{equation}
while for the $\rho_{03}\equiv \rho$ field
\begin{equation}m_{\rho}^{2}\rho = \sum\limits_{B} g_{\rho B}\tau_{3B}n_{B},
\end{equation}
being $\tau_{3B}={\rm diag}(1/2,-1/2)$ the isospin 3-rd component matrix operator.  Finally for the $\sigma$ field, 
\small
\begin{equation}
\begin{aligned}
& m_{\sigma}^{2}\sigma= -\frac{d\mathcal{U}(\sigma)}{d\sigma} + \frac{1}{2\pi^{2}}\sum\limits_{B}g_{\sigma B}{m^*_B}^{3}\left[ t_{B}\sqrt{1+t_{B}^{2}} \right.-\\
& {\rm ln}\left.\left( t_{B}+\sqrt{1+t_{B}^{2}} \right)\right].\\
\end{aligned}
\label{sigma}
\end{equation}
\normalsize
where we have defined $t_B=k_{F,B}/m^*_{B}$. We can also write the chemical potential for baryons as $\mu_{B} = \sqrt{k_{F,B}^{2}+{m^*_B}^{2}} + g_{\omega B}\omega + g_{\rho B}\tau_{3B}\rho$ while for leptons $\mu_{l} = \sqrt{k_{F,l}^{2}+m_{l}^{2}}$. Additionally, imposing conditions for  electrical charge neutrality $n_e=n_p$ and $\beta$-equilibrium $\mu_n=\mu_p+\mu_e$ (no neutrinos are trapped) we can obtain the solution for the mesonic fields, provided a $n_b$ value is set. 
This allows to obtain the EoS from contribution of all particle species. From this, the total energy density and pressure to describe the interior of the old NS can be written as a sum of hadronic (Had) and leptonic (l) terms, $\varepsilon = \varepsilon_{\rm Had} + \varepsilon_{l}$ and $P = P_{\rm Had} + P_{l}$, respectively. More explicitly,
\small
\begin{equation}
\begin{aligned}
\varepsilon = {} 
 & \frac{1}{8\pi^{2}}\sum\limits_{i=B,l}{m^*}_i^{4}\left[ (2t_{i}^{2}+1)t_{i}\sqrt{1+t_{i}^{2}}-ln\left( t_{i}+\sqrt{1+t_{i}^{2}} \right) \right]\\
 & + \mathcal{U}(\sigma) + \frac{1}{2}m_{\sigma}^{2}\sigma^{2} + \frac{1}{2}m_{\omega}^{2}\omega^{2} +\frac{1}{2}m_{\rho}^{2}\rho^{2},\\ 
\end{aligned}
\label{ehad}
\end{equation}
\normalsize
and 
\begin{equation}
P=-\sum\limits_{i=B,l}\varepsilon_{i}+\sum\limits_{i=B,l} n_i\mu_{i}.
\end{equation}
We use $t_{l}=k_{F,l}/m_l$ and ${m^*}_l=m_l=m_e$.

Let us note that although a proper treatment of the hadronic (quark) system would require using the framework of Quantum Chromodynamics (QCD) at intermediate energies, this calculation is, in  practice, technically infeasible. It is for this reason that phenomenologically different approaches using effective field theories with baryonic (mesonic) and more fundamental quark (gluon) degrees of freedom have been exploited in order to determine the EoS of both forms of matter consistently \cite{pnjl} and explore their thermodynamical conditions of stability. In this spirit, to describe the deconfined quark matter phase we use the MIT Bag model \cite{mit} where gluon fields are  effectively considered through a vacuum pressure, $B$, including interactions through the strong coupling constant $\alpha_S$ \cite{alford}. The perturbative QCD parameter $\alpha_S$ characterizes the degree of the quark interaction correction, with $\alpha_S=0$ corresponding to no QCD corrections (Fermi gas approximation). For our purposes we will explore values $\alpha_{S}=0.4,0.5$ in line with previous works \cite{Iida}. In addition, these selected values are also included in the allowed range arising from gravitational wave and astrophysical constraints quoted in  \cite{alpha1,alpha2}. We use this model as it provides a tractable and meaningful way to describe a hypothetical more fundamental configuration of matter although other more refined approaches exist \cite{nambu}. Regarding this aspect we expect no dramatic modification of the results we find.

\begin{figure}[t]
\begin{center}
\includegraphics [angle=0,scale=0.85] {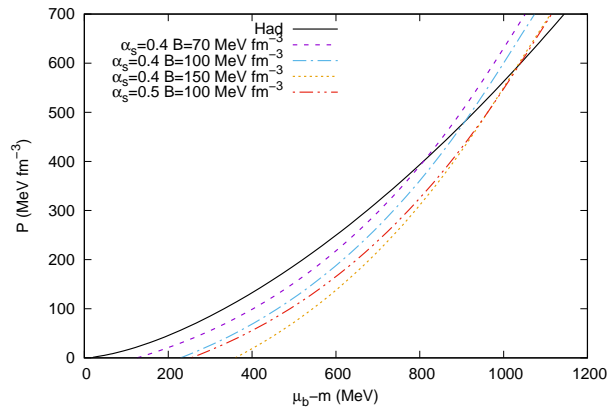}
\caption{Pressure as a function of the baryonic chemical potential for hadronic matter (Had) and deconfined $ud$ matter. We use $\alpha_{s}=0.4$, $0.5$, $B=70, 100, 150$ $\rm MeV\, fm^{-3}$.}
\label{fig1}
\end{center}
\end{figure}

For a cold $uds$ quark system, the thermodynamical potential $\Omega_{q}$ for each light flavour with mass $m_q$ and chemical potential $\mu_q$ ($q=u,d,s$), has a general form 
\small
\begin{equation}
\begin{aligned}\label{PotencialQuark}
\Omega_{q} = & -\frac{1}{4\pi^{2}} \left[ \mu_{q}p_{\rm F,q}\left( \mu_{q}^{2}-\frac{5}{2}m_{q}^{2} \right) 
 +\frac{3}{2}m_{q}^{4}\,\rm{ln}\left( \frac{\mu_{q}+p_{\rm F,q}}{m_{q}} \right) \right] & \\
 & + \frac{\alpha_{S}}{2\pi^{3}}\Bigg\{ 3\left[ \mu_{q}p_{\rm F,q}-m_{q}^{2}\,\rm{ln}\left( \frac{\mu_{q}+p_{\rm F,q}}{m_{q}} \right) \right]^{2}-2p^4_{\rm F,q}) \Bigg\},
\end{aligned}
\end{equation}
\normalsize
where $p_{\rm F,q}=(\mu_{q}^{2}-m_{q}^{2})^{1/2}$. Here we use the approximation $m_{u}=m_{d}\equiv0$ and $m_s=150$ MeV. Thus for massless quarks $\Omega_{q}$ adopts the simplified form $\Omega_{q}=-\frac{\mu_{q}^{4}}{4\pi^{2}}\left( 1-\frac{2\alpha_{S}}{\pi} \right)$. The quark number density is obtained as $n_{q}=-\frac{\partial\Omega_{q}}{\partial\mu_{q}}$. 

Finally, the energy density for $uds$ matter includes the contribution from the effective bag constant $B$ under the form $\varepsilon_{\rm Quark} = \sum\limits_{q}\left( \Omega_{q}+\mu_{q}n_{q} \right) + B$ or, more explicitly,
\small
\begin{equation}
\begin{aligned}
\varepsilon_{\rm Quark} =& \frac{3}{4\pi^{2}}\left( 1-\frac{2\alpha_{s}}{\pi} \right)(\mu_{u}^{4}+\mu_{d}^{4}) + \frac{3}{8\pi^{2}}m_{s}^{4}\left[ x_{s}\eta_{s}(2x_{s}^{2}+1)  \right. \\
& -\left. {\rm ln}(x_{s}+\eta_{s}) \right] -\frac{\alpha_{s}}{2\pi^{3}}m_{s}^{4}\left\{ 2x_{s}^{2}(x_{s}^{2}+2\eta_{s}^{2})-3\left[ x_{s}\eta_{s}\right. \right. & \\
& +\left. \left. {\rm ln}(x_{s}+\eta_{s}) \right]^{2} \right\}+B, 
\end{aligned}
\end{equation}

\normalsize
where $x_{s}=\sqrt{\mu_{s}^{2}-m_{s}^{2}}/m_{s}$, $\eta_{s}=\sqrt{1+x_{s}^{2}}$. Note that an additional lepton component $\varepsilon_{l}$ must be present in the total $\varepsilon$ contribution, analogous to that in Eq. (\ref{ehad}). For the quark-gluon pressure we have
\begin{equation}\label{PresionQuark}
P_{\rm Quark} = -\sum\limits_{q}\Omega_{q}-B
\end{equation}
and the total pressure is $P = P_{\rm Quark}+ P_{l}$. As before, the conservation of baryonic charge, weak equilibrium and the additional constraint of electric charge neutrality in NS matter can be expressed for the quark phases as $n_b=\frac{1}{3} \sum_q n_q$, $\mu_d=\mu_u+\mu_e$, $\mu_s=\mu_d$ and $\frac{2}{3}n_u=\frac{1}{3}(n_d+n_s)+n_e$. 
In this model the actual first order phase transition comes determined by the thermodynamical conditions of the cold NS interior \cite{limit}. For the multicomponent system under scrutiny the Gibbs criteria imposes the equality of baryonic chemical potentials at a given pressure $P_{0}$ where equilibrium holds,
\begin{eqnarray}
&\mu_{b}(P_{0})\Big|_{\textrm{Had}} = \mu_{b}(P_{0})\Big|_{\textrm{Quark}},&
\end{eqnarray}
and $\mu_{b} = \frac{\varepsilon+P}{n_{b}}$ in the cold system. 

Let us remark here that, on general grounds, the nature of the transition between the Hadron-Resonance
Gas (HRG) and Quark-Gluon Plasma (QGP) phases of nuclear matter can be presented in terms of the phase
diagram of QCD as a function of temperature T and baryon chemical potential $\mu_b$. At large T and small  $\mu_b$, the transition is expected on the basis of lattice calculations to be a rapid crossover, whereas it is naturally expected that the phase transition
along the  $\mu_b$-axis (with $T = 0$) is an actual first-order phase transition \cite{raja, buba}. Somewhere along
the phase transition line the point at which this occurs is commonly known as the QCD critical point  that has not yet been experimentally confirmed nor predicted with theoretical certainty \cite{cp}. 


In our stellar scenario DM with $m_\chi\gtrsim 10$ GeV and cross sections $\sigma_{\chi N}$ in the currently allowed phase space can thermalize and display a radial distribution inside the NS,  $\rho_{\chi,*}(r)$, built on top of that of ordinary baryonic matter. The ratio of gravitational to thermal energy makes DM to follow a Gaussian distribution $\rho_{\chi,*}(r)\sim e^{-(r/r_{\rm th})^2}$ with a radius $r_{\rm th}= \sqrt{\frac{3 k_B T}{2 \pi G \rho_B m_{\chi}}}$. This central region is where most DM annihilations will take place. We consider that the inner NS core has a baryonic mass density around three times that of nuclear saturation density, $\rho_B\sim 7\times 10^{14}$ $\rm g/cm^3$, and an internal temperature $T\sim 10^{5}$ K at a given galactic location with a corresponding ambient DM density, $\rho_\chi$. Typical values in the solar neighbourhood are $\rho_{\chi,\rm local}= (0.39 \pm 0.03)$ $\rm GeV/cm^3$ \cite{catena} although NSs are found closer to the galactic centre where they can be higher by a factor of $\sim$100. The NS captures DM, up to factors of order unity, at a rate $C_\chi$ \citep{gould,lavallaz} 
\begin{equation}
C_{\chi}\simeq 1.8 \times 10^{23} \left(\frac{100\, \rm GeV}{m_{\chi}}\right)\left(\frac{\rho_{\chi}}{\rho_{\chi,\rm local}}\right) f_{\chi,N}\,\,\rm s^{-1}.
\label{captu}
\end{equation}
\normalsize
 $f_{\chi,N}$ denotes a phenomenological factor dealing with the opacity of stellar matter as it depends on the ratio of the leading contribution of $\chi N$ scattering cross section $\sigma_{\chi N}$ to the minimum geometrical cross section defined as $\sigma_0\sim \frac{m_B}{M_*}R_*^2\sim 10^{-45}\rm cm^2$. Thus this factor saturates to unity $f_{\chi,N}\sim 1$ if  $\sigma_{\chi N}\gtrsim {\sigma_0}$,  whereas it decreases the capture rate otherwise. As current experimental efforts foresee sensitivities below $\sigma_0$ for some $m_\chi$ windows, this would imply lower amounts of DM inside the NS by this same factor. However this aspect is not critical as a tiny content of DM can induce important changes in the NS EoS, as we will explain.

The DM particle population number inside the star, $N_\chi$, will  not only depend on the capture rate $C_\chi$ but also on the self-annihilation rate, $C_a$, in our scenario. This latter is model-dependent but we can estimate it to be $C_a\sim \langle \sigma_a v \rangle /r_{\rm th}^3$. Numerically,
\footnotesize
\begin{equation}
C_a=1.1 \times 10^{-30} \left( \frac{ \langle \sigma_a v \rangle} {3\times 10^{-26}\,\rm cm^3 s^{-1}}\right) \left(\frac{10^5\,\rm K}{T} \frac{m_\chi}{100\,\rm GeV} \right)^{3/2} 
\,\rm s^{-1}.
\end{equation}
\normalsize
Therefore $N_{\chi}$ can be obtained as a function of time $t$ by solving the differential equation for the NS  core $\frac{dN_{\chi}}{dt}=C_{\chi}-C_a N^2_{\chi}$ considering the two competing processes of capture and self-annihilation. The solution can be written as
\begin{equation}
N_\chi(t)=\sqrt{\frac{C_\chi}{C_a}} \rm \,tanh \left[\frac{t}{\tau}+\gamma (N_{\chi,0})\right],
\label{Nx}
\end{equation}
where $N_{\chi,0}$ is the DM population in the final stage of its progenitor phase and $\gamma (N_{\chi,0})={\rm tanh}^{-1} \left(\sqrt{\frac{C_\chi}{C_a}}N_{\chi,0}\right)$ with $\tau^{-1}=\sqrt{C_\chi C_a}$. For times large enough so that the system has reached the steady state, $t\gg \tau$, the asymptotic population is given by $N_\chi(t_{\infty})\simeq \sqrt{\frac{C_\chi}{C_a}}$.\\
Under such conditions, we model the average energy release from annihilation processes as obtained from the spectrum of the reaction $\chi {\bar \chi} \rightarrow q {\bar q} \rightarrow N \gamma$. Although other channels are indeed possible \cite{cirelli}, we restrict our modeling to this one as the mechanism presented here will not be dramatically altered. The spectrum $\frac{dN_\gamma}{dE}$ provides the average energy release as $\langle E \rangle = \int_{0}^{m_{\chi}} E\frac{dN_\gamma}{dE}\,dE$. Note that the upper limit in the integral is due to the fact that the center of mass energy is $E_{\rm CM}=\sqrt{s}=2 m_\chi$ being $s$ the Mandelstam variable and each quark (antiquark) carries a maximum value $E_{\rm q}=\sqrt{s}/2=m_\chi$.  We have used the PYTHIA package \cite{pythia} to obtain that as an indication,  for light quarks approximately $67\%$ of the DM mass $m_\chi$ is deposited in the medium in the form of photons. Heavier quarks, alternative channels or even corrections due to energy loss and final state kinematics from the most energetic photons would indeed change the size of the energy injection. Since the energetics of the microscopic particle physics event in the very dense medium inside the NS core with $\sim10^{34}$ $\rm nucleons/cm^3$ is rather complex this treatment constitutes just an approximate description. However, we do not expect it dramatically alters the nucleation mechanism itself as exposed here.

Typically, the photons produced will be obtained from different highly energetic hadronic as well as electromagnetic showers involving inelastic interactions with the dense medium. In this way, the hadrons with a cross-section $\sigma_{\rm Had}$, which is largely uncertain but can be roughly estimated to be $\sigma_{\rm Had}\sim 1$  $\rm fm^2$ will produce showers with typical sizes depending on the baryonic density in the NS core. Taking a density $n_b\sim (3-5)n_0$ and typical energies $E\sim 1-10^5$ GeV for each event we obtain a size $X_{\rm Had}\sim \frac{1}{n_b \sigma_{\rm Had}} \rm Log \left( \frac{E}{10\,\rm MeV} \right) \lesssim 10$ fm for the hadronic showers while $X_{\rm EM}\sim 1$ fm for the electromagnetic showers \cite{graham}. The injected energy is mainly contained into the bubble region with radius $R$ as long as $X_{\rm Had} \lesssim 2 R$.
\begin{figure}[t]
\begin{minipage}[c][11cm][t]{.5\textwidth}
  \centering
  \includegraphics[scale=0.85]{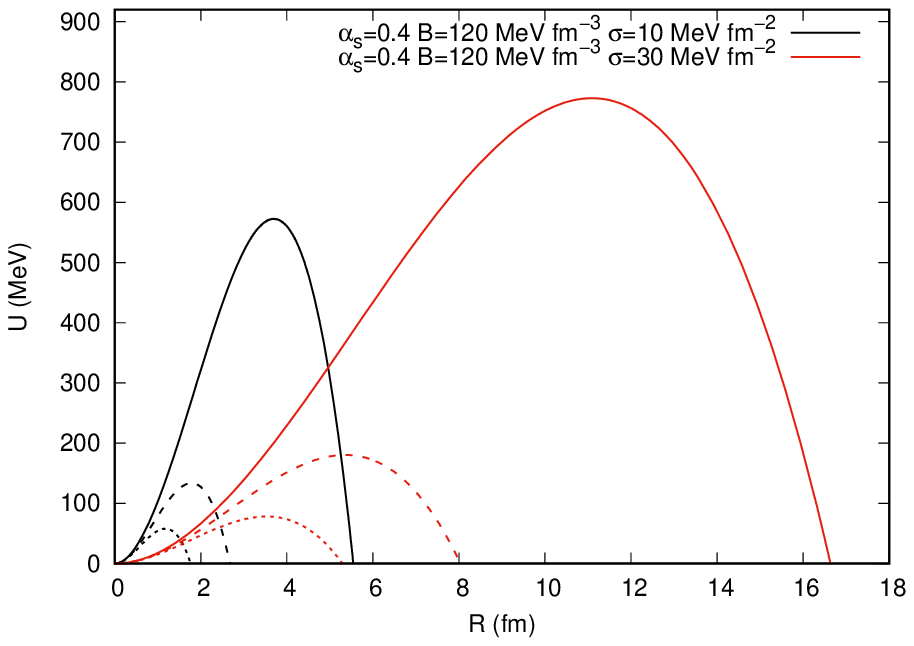}
  \includegraphics[scale=0.85]{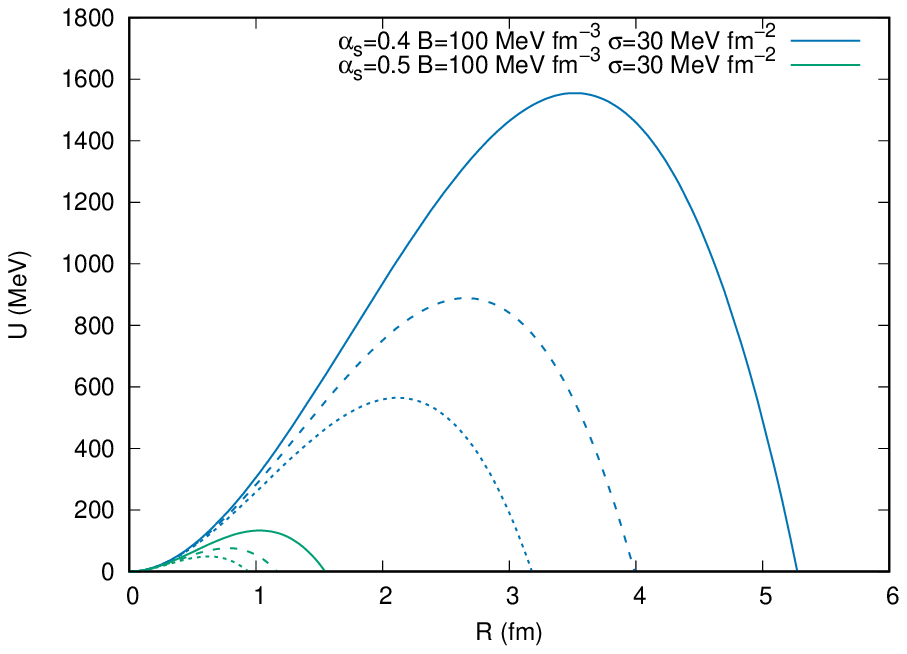}
\end{minipage}
 \caption{Potential barrier energy as a function of radius for different combinations of $(\alpha_S,B,\sigma$). Solid, dashed and dotted lines depict $P=P_0+32$ MeV, $P=P_0+42$ MeV, $P=P_0+52$ MeV for $B =120$ $\rm MeV/fm^3$ (upper panel) and  $B =100$ $\rm MeV/fm^3$ (lower panel). Red curves on the upper panel have been scaled a factor $1/20$ to improve readability. See text for details.}
 \label{fig2}
\end{figure}

The global picture is thus that of a central DM annihilation volume, where both types of matter coexist, and baryonic matter is subject to  steady state energy injection from the quoted DM reactions. To explore whether this scenario could lead to induced quark bubble nucleation we use the relativistic Lifshitz-Kagan theory \cite{lk}. We aim to describe  the microscopics of a locally induced phase transition from an effective (metastable) hadronic phase to a more fundamental  deconfined state. In this formalism the relativistic lagrangian that describes the formation of a fluctuation i.e. a spherical bubble of mass $M$ and radius $R$ is given by 
\begin{equation}\label{LagrangianoBurbuja}
\small
\mathcal{L}(R,\dot{R}) = -M(R)\sqrt{1-\dot{R}^{2}} + M(R)-U(R),
\end{equation}
\normalsize
where $\dot{R}=dR/dt$ is the radial growth rate 
and $U(R)$ is an effective potential depending on the thermodynamical conditions of the medium
\begin{equation}\label{PotencialBurbuja}
U(R) = \frac{4}{3}\pi R^{3}n_{\rm Quark}(\mu_{\rm Quark}-\mu_{\rm Had})+4\pi\sigma R^{2}.
\end{equation}
We  label $\mu_{\rm Had}$ ( $\mu_{\rm Quark}$) as the chemical potentials of the hadronic (quark) phases of matter at fixed pressure value, $P$. $n_{\rm Quark}$ is the number density of the quark phase and $\sigma$ is the surface tension among the two phases. Contributions from volume as well as surface terms are the most energetically relevant although more corrections can be included \cite{olesen}. As written, this expression may be familiar to the reader as it also used in terrestrial experiments for DM searches such as PICO \cite{pico} at SNOLAB, with an analogous strategy based on detecting $\chi$ scattering events in bubble chambers filled with overheated liquids at much smaller densities.

When nucleated, QM bubbles can be characterized by a critical radius for stability, 
\begin{equation}
R_{c} = \frac{3\sigma}{n_{\rm Quark}(\mu_{\rm Quark}-\mu_{\rm Had})},
\end{equation}
fulfilling the relation $U(R_{c})=0$ such that for $R>R_c$, the bubble is energetically stable. The potential barrier maximum height to be tunneled is 
\begin{equation}
U_{\rm max} = \frac{16}{27}\pi\sigma R_{c}^{2}.
\end{equation}
In order to drive further changes at the macroscopic level QM bubbles must have the capability to last sufficiently and, in any case, longer than the time scale for stellar dynamical collapse $\tau_D \sim \sqrt{2R_*^3/2 GM_*}\sim 3\times 10^{-5}$ s.

Making a transformation to the phase space canonical variables $q,p$ one can obtain the Hamiltonian associated to the lagrangian in Eq.(\ref{LagrangianoBurbuja}) as $\mathcal{H}(q,p) = p\,\dot{q} - \mathcal{L}$. As explained in  \cite{Iida} by taking the \textit{WKB} approximation \cite{WKB}, one can compute the energy of the ground state, $E_0$. The associated oscillation frequency $\nu^{-1}_0=\frac{dI}{dE}|_{E=E_0}$ is obtained \cite{Iida} using the expression 
\begin{equation}
I(E) ={2}\int_{0}^{R_{-}} \sqrt{\left[ 2M(R)+E-U(R) \right] \left[ U(R)-E \right]}\,dR,
\end{equation}
being $R_{-}$ the smaller turn around radius. The action under the potential barrier
\small
\begin{equation}\label{AccionBurbuja}
A(E) = {2}\int_{R_{-}}^{R_{+}} \sqrt{\left[ 2M(R)+E-U(R) \right] \left[ U(R)-E \right]}\,dR.
\end{equation}
\normalsize
determines the tunneling probability $p_{0} = \rm exp\left( -\frac{A(E)}{\hbar} \right)$.

In this system, the average energy per quark  injected into the nucleon bag from DM self-annihilation events necessary to create a stable spherical droplet of size $\sim R_c$ can be approximated by  $E_{\rm inj}\sim \frac{\langle E \rangle}{n_b 4 \pi R^3_c }$. We assume that the energetic shower is contained into the bubble, however, if this is not the case a decreasing factor $\xi\sim (2 \rm R/X_{\rm Had})$ must be included. Thus the photon yields from DM self-annihilation act effectively to raise the energy level of otherwise confined quarks. 

The bubble formation time is obtained as $\tau^{-1}=\nu_0 p_0$. It is important to notice that the nucleation process may most probably happen over the whole DM thermal volume $\sim r^3_{\rm th}$. Therefore a number of nucleation centers are available for this process and can be estimated as $N_{C} \sim (\frac{r_{\rm th}}{R_c})^3$. In such a case the corrected nucleation time is given by  $\tau_N=\tau/N_{C}$.  As explained in the introduction section of this contribution, the nucleation of quark droplets relying on standard astrophysical mechanisms is highly suppressed. This means, in practice, that nucleation times may be much larger than stellar lifetimes. As we will explain below, as a result of an extra energy injection from DM self-annihilation the hadronic system can be driven into an excited configuration allowing the formation of stable quark droplets more easily. 

In our description of the cold nuclear system we have not considered the possibility of the existence of a fraction or paired nucleons. Typically for the high densities and low temperatures in the NS core the paired nucleon fraction is less than $\sim 10\%$. At supranuclear densities the pairing gaps, $\Delta$, for neutrons ($^3P_2$–$
^3F_2$ channels) and for protons ($^1S_0$ channel) have been estimated yielding sizes up to a few MeV. Instead,  for quarks they may be as large as a hundred MeV, see for example \cite{sedra} for a review. In brief, the density of paired nucleons depends on this gap \cite{n}  as  well as the quantity $\sqrt{(E_N(k,m^*_N)-\mu^*_N)^2+\Delta^2(k)}$ involving the nucleon single-particle energy $E_N$, effective mass $m^*_N$ and effective chemical potential $\mu^*_N$ for a fixed momentum $k<k_{F,N}$. The existence of such a gapped fraction  adds an extra amount of energy  ($E_{\rm BCS\, pair\, break}>2\Delta (k_{F,N}$)) that must be overcome in order to break the correlated pair as dictated by the BCS theory \cite{bcs}. In this sense, the possible existence of a paired fraction tends to suppress the nucleation process. However, it is worth noting that, safely, for nucleons $\Delta (k_{F,N})/m_\chi \ll 1$ for typical NS core densities and most of the $m_\chi$ range explored in our study and similar applies for quarks. Thus we expect that the possible corrections from kinematically blocked states in the final phase space or paired components will not have a dramatic impact and the scattered fermions will have energies above the Fermi level for DM masses larger than a few GeV. However, this fact must be carefrully reconsidered for $m_\chi\lesssim 1$ GeV as this could be a competing effect quenching the efficiency of nucleation. For simplicity we have neglected this  possibility leaving it for a future contribution. 

\section{Results}
In this section we present our results. We have used arbitrary values $B\in[70,150]$ $\rm MeV\, fm^{-3}$ as usually done in the literature, however recent calculations \cite{dexheimer} restrict stable solutions of $T=0$ $uds$ QM to a window $60 \lesssim B  \lesssim 80$ $\rm MeV\, fm^{-3}$. In addition, if confirmed, quark stars having masses beyond $\sim 2M_\odot$ limit would imply even smaller $B$ values \cite{alpha1}. In addition values of the surface tension are poorly known. This quantity is relevant for bubble nucleation in different already studied environments, for example quark matter in supernovae \cite{snova} and neutron star mergers \cite{merger}. It is also relevant for a possible quark-hadron mixed phase in the interior of neutron stars \cite{glend}. Some available calculations  have estimated its value using different complementary approaches such as  Nambu-Jona-Lasinio models, quark-meson models \cite{njl} or chiral models  \cite{sigmafraga} where in the latter maximum values point towards  $\sigma \sim 15$ $\rm MeV\, fm^{-2}$. For our purposes we will consider $\sigma \in[10,30]$ $\rm MeV\, fm^{-2}$.

 In Fig.(\ref{fig1}) we plot pressure as a function of the baryonic chemical potential $\mu_b-m_B$ for hadronic matter (Had) and deconfined $ud$ matter. We use $\alpha_{S}=0.4$, $0.5$, $B=70, 100, 150$ $\rm MeV\, fm^{-3}$ and $\sigma=30$ $\rm MeV\, fm^{-2}$. As the baryonic chemical potential (density) grows the softer quark matter EoS overpasses that of hadronic matter at a transition point i.e. the crossing of both curves. The higher the $B$ ($\alpha_S$) the higher the chemical potential associated to the transition density. The transition pressures and baryonic densities for $\alpha_S = 0.4$ are $P_0 = 588, 483, 401$ $\rm MeV\,fm^{-3}$ and $n_b = 0.92,0.87,\,0.78$ $\rm fm^{-3}$ for  $B=150,100,70$ $\rm MeV fm^{-3}$, respectively. These results do not depend on $\sigma$. If instead we use $\alpha_{S}=0.5$ and $B =100$ $\rm MeV\,fm^{-3}$, we find $P_0=600$ $\rm MeV\, fm^{-3}$, $n_{\rm Had}=0.95$ $\rm fm^{-3}$, $n_{\rm Quark} = 1.80$ $\rm fm^{-3}$.

In Fig.(\ref{fig2}) we  plot the potential barrier $U(R)$ as a function of radius for different parameter sets. Solid, dashed and dotted lines depict $P=P_0+32$ MeV, $P=P_0+42$ MeV, $P=P_0+52$ MeV for each case. In the upper panel we fix $\alpha_{S}=0.4$, $B=120$ $\rm MeV\, fm^{-3}$, $\sigma=10$ $\rm MeV \,fm^{-2}$ (black curves) and $\sigma=30$ $\rm MeV \,fm^{-2}$ (red curves). The latter case (red lines) has been scaled a factor $1/20$ in order to numerically compare on the same axis. The pressure at the transition point is $P_0=525$ $\rm MeV \,fm^{-3}$ while the density is $n_{\rm Had}=0.88$ $\rm fm^{-3}$, $n_{\rm Quark} = 1.58$ $\rm fm^{-3}$. We can see that for higher $\sigma$ values, larger bubbles are required for stability and, at the same time, more energetic barriers form. Instead, for increasing pressure, smaller bubbles can survive. In the lower panel we fix $B=100$ $\rm MeV \,fm^{-3}$, $\sigma=30$ $\rm MeV \,fm^{-2}$, $\alpha_{S}=0.4$ (blue curves) and $\alpha_{s}=0.5$ (green curves). We can clearly see that for larger $\alpha_{S}$ values i.e. including less strong corrections, smaller bubbles are predicted.

\begin{figure}[t]
\begin{center}
\includegraphics [angle=0,scale=0.85] {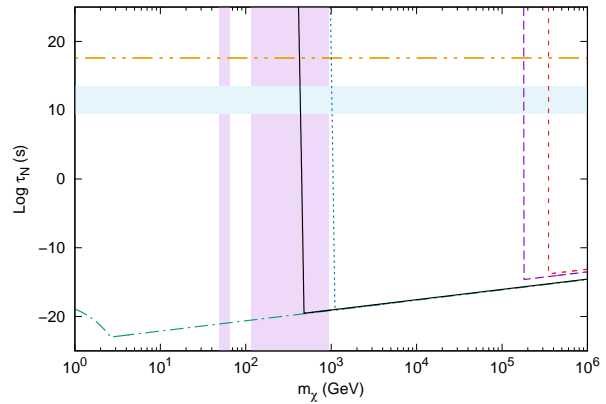}
\caption{Nucleation time as a function of the DM particle mass for different cases. Coloured bands depict old NS ages (blue horizontal) and SUSY favoured regions (pink vertical). The Universe lifetime $\tau_U\sim 10^{17.6}$ s is also shown. See text for details.}
\label{fig4}
\end{center}
\end{figure}

In Fig.(\ref{fig4}) we  plot the logarithm (base 10) of the nucleation time as a function of the DM particle mass for different parameter sets used. The amount of DM for each $m_\chi$ value is obtained from the  asymptotic population $N_\chi(t_\infty)$ ranging from $N_\chi\sim 10^{30}$ for $m_\chi=1$ GeV to $N_\chi\sim 10^{23}$ for $m_\chi=10^6$ GeV. These values are in all cases below the critical value given by their fermionic or bosonic nature, see \cite{perez, ch}. In addition, given the spread of the hadronic/quark parameters scanned, the pressure values considered for evaluation are selected above their corresponding $P_0$ for each case.

For the first set we fix $\alpha_{S}=0.4$, $B=70$ $\rm MeV\, fm^{-3}$, $\sigma=10$ $\rm MeV\, fm^{-2}$ and select $P=406$ $\rm MeV \,fm^{-3}$ (long dashed). For the second set (same $\alpha_{S}$ as before) $B=120$ $\rm MeV \,fm^{-3}$, $\sigma=30$ $\rm MeV\, fm^{-2}$ with  $P=535$ $\rm MeV \,fm^{-3}$ (short dashed), and third set $\sigma=10$ $\rm MeV \,fm^{-2}$ (solid). For the fourth one $B=100$ $\rm MeV \,fm^{-3}$, $\sigma=30$ $\rm MeV \,fm^{-2}$ with $P=515$ $\rm MeV\, fm^{-3}$ (dotted) and, finally, the fith set is the same as the latter case but using $\alpha_{S}=0.5$ (dotdashed). 

We plot additional coloured regions with constraints coming from NS ages typically measured (blue horizontal), see Table I in \cite{cool} with $\tau_{NS}\sim 10^{9.5}-10^{13.5}$ s. They can even reach $\sim 4.9$ Gyr as in PSR J0437–4715 \cite{johns}, close to the upper limit provided by the Universe lifetime (yellow doble dash-dotted line) with $\tau_U\sim 10^{17.6}$ s. We also plot the regions (pink vertical) denoting the preferred four typical SUSY models, CMSSM, NUHM1, NUHM2, pMSSM10 which integrates constraints set by ATLAS Run 1 \cite{pdg}.

We can see that nucleation times are critically dependent on the $B,\alpha_S$ and $\sigma$ values of the quark phase. The thermodynamical conditions (central pressure) produce a moderate impact on the results. Increasing the bubble critical radius i.e. $\sigma$, produces a high energetic cost. The energy injection obtained from the self-annihilating $\chi$ pair signals the kink  where probability of nucleation $p_0\sim 1$. The results obtained for the $B=70$ $\rm MeV \,fm^{-3}$ set properly belong to the quoted stability window of QM as obtained in earlier works \cite{dexheimer}. This case requires a $m_\chi\gtrsim 4\times 10^5$ GeV to induce nucleation in the central thermal volume inside the NS. In this scenario, light DM candidates ($m_\chi <10$ GeV) would less favour  such a conversion in ordinary NSs. Baryonic densities at the transition point found for each case are indeed in the previously estimated interval  $\sim (3-4)n_0$. Note that additional corrections due to final state limitations of the energy injected as well as hadronic cascade sizes versus bubble radius will  correct the energy efficiency by factors $\mathcal{O}(1)$. Thermal energy loses are found to be negligible in a cold system. In addition, other basic channels
into which DM particles may annihilate could happen, including heavy quarks, leptons, weak or Higgs bosons \cite{annih}. Besides, a direct neutrino/antineutrino  annihilation channel or neutrinos originating  from the decays of the particles produced in the annihilation could take place. This could modify effectively the energy injected into the central hadronic region altering the efficiency of the process \cite{lineros}. The hierarchy
problem, also framed as requiring naturalness or the absence of fine tuning, prefers WIMP masses below about a few TeV, which is considerably more constraining than the WIMP coincidence which essentially allows particles from a few GeV to about 100 TeV \cite{griest}. Nevertheless in our analysis we have considered maximum values of $m_\chi \lesssim 10^3$ TeV in consistency with the wimp-like scenario depicted here and with capture rates in accordance with a typical weakly interacting DM candidate  being accreted by the NS, see results from CTA \cite{cta} or MAGIC \cite{magic}. If more massive DM candidates were considered additional corrections to several quantities e. g. capture rates, effective hadronic potentials, corrections from final phase space multiple channels or excited states of quark degrees of freedom should be considered.

A remark is due at this point regarding the feasibility of the percolation transition in the macroscopic stellar object. The mechanism quoted for nucleation in the  relativistic Lifshitz-Kagan theory involves matter undergoing this process is at a metastable state i.e. at pressures larger than the transition pressure $P_0$. Given an object with mass $M_*$ and radius $R_*$ it may happen that displays either pure hadronic or nuclear-quark hybrid mixed nature. As we mentioned in the introduction section we have focused on the nucleation mechanism for matter in NS cores under the hadronic state in consistency with the DM-nucleon scattering cross sections used in the capture rate in Eq. (\ref{captu}). Additional corrections would arise from a hybrid NS or even a quark star and will be treated elsewere.

In the hadronic stellar object the existence of such a macroscopic percolated region is expected to convert the full NS as already explained in \cite{bereziani} producing an energetic $10^{51}-10^{53}$ erg gamma ray burst (GRB). On the other hand for some of the parameter sets explored in our work (see dot-dashed line curve in Fig. (3)) this mechanism would induce a rapid nucleation that, if progressing to macroscopic, seems to be hardly reconcilable with the frequency of very short GRBs observed \cite{daigne}. The correspondance of the $m_\chi$ to the probability of nucleation of quark droplets seems remarkably sensitive under this mechanism. At a microscopic level, a dynamical study of the droplet boundary once formed has not been studied in detail yet. If the full NS converts to a more compact QM star this exotic type of matter would be ejected leaving an imprint on the cosmic rays or scintillation patterns \cite{paulucci, kumiko,pen}. Boundary conditions for isolated clusters have been somewhat explored in \cite{gilson} where, arising from the strange quark content, the name of {\it strangelet} is coined. If a strangelet is not in flavour equilibrium, it can decay via weak semileptonic decays, weak radiative decays and electron capture. Other modes of decay reduce the baryon number instead. Further work on this is under progress and will be reported elsewhere.

\section {Conclusions} 

We have studied the nucleation of quark droplets in the NS core facilitated by the energy injection due to a component of self-annihilating DM present inside neutron stars. We find that under the effective field theoretical description of the hadronic and quark phases, the latter using a MIT bag model, the dark matter candidate mass highly influences the nucleation time $\tau_N$. Depending on the central pressure (density) conditions inside the star nucleation times may span 40 orders of magnitude. Within this scenario light dark matter is less favoured to produce such a percolation phase transition inside the core central region. However, for parameter sets within a window of energetic stability for stellar mass-radius values of $m_\chi\gtrsim 4 \times 10^5$ GeV are found capable of nucleating  the macroscopic thermal DM core during the NS lifetime and drive a conversion into a more compact  star. Emission of radiation (GRB) or chunks of matter (cosmic rays) is to be expected along with gravitational waves. To determine the temporal sequence of the multi-messenger signal and magnitude of the effects presented here, detailed calculations of the central EoS instability are needed and are left for future work.

\section {Acknowledgments}
We would like to thank S. Heinemeyer, C. Kouvaris  for useful discussions. This work has been supported by Junta de Castilla y Le\'on SA083P17. We also thank the support of the Spanish Red Consolider MultiDark FPA2017-90566-REDC, PHAROS Cost action and University of Salamanca. 

\end{document}